\documentclass[10pt,a4paper]{article}
\usepackage{latexsym}
\usepackage{lscape}
\usepackage{graphicx}
\usepackage{hyperref}
\usepackage{microtype}
\usepackage{eurosym}
\usepackage{authblk}

\title{A review of approaches to the value of privacy}
\author[1]{Miltiades E. Anagnostou}
\affil[1]{School of Electrical and Computer Engineering, National Technical University of Athens,
 Zografou 15780, Greece\\ email:\href{mailto:miltos@central.ntua.gr}{miltos@central.ntua.gr}}
\author[2]{Maria A. Lambrou}
\affil[2]{Dept. of Shipping Trade and Transport, University of the Aegean,
Chios 82132, Greece\\
email:\href{mailto:mlambrou@aegean.gr}{mlambrou@aegean.gr}
}

\begin{document}



\maketitle

\begin{abstract}
In this paper we examine information privacy   from a value-centered angle. We review and compare different personal data valuation approaches. We consider the value of personal data now and in the near  future, and we find that the enduring part of personal data will soon become common knowledge and that their price and value will drop substantially. Therefore the sector that is based on personal data extraction will need to focus on new ways of monetization and on the of the dynamic part of personal information.

\end{abstract}

\section{Introduction}
\label{sect:introduction}

In a classical definition  privacy is a special case of a {\em right to be let alone} \cite{warren1890right}. In the Random House Unabridged Dictionary privacy is {\em the state of being free from unwanted or undue intrusion or disturbance in one's private life or affairs; freedom to be let alone}.
Tavani  distinguishes between three types of privacy \cite{tavani2011ethics}: (a)~Accessibility (related with physical space), (b)~decisional (right to make decisions freely), and (c)~informational privacy, the latter being defined as {\em control over the flow of one's personal information}.
According to EU Regulation 2016/679 {\em the protection of natural persons in relation to the processing of personal data is a fundamental right}, where {\em ``personal data'' means any information relating to an identified or identifiable natural person}. Of particular importance are the {\em sensitive personal data},  which are {\em personal data revealing racial or ethnic origin, political opinions, religious or philosophical beliefs, trade-union membership, and data concerning health or sex life}.


In this paper, we explore the privacy protection problem, both for online social networks and generic e-service systems, from a value-centered angle. 
 
\section{Value is created by using personal data}


\subsection{Value of personal data and value of services}
There is a realm of apps and services, which includes social networking, that are provided for ``free,'' subject to user consent to privacy related terms. 
Service providers record and analyze  user behavior for their own financial support, which is achieved mostly by targeted advertising. Users, who are aware of these practices may want to protect their personal data, sell them for a satisfactory price, or just wonder whether ``free'' services are a fair compensation.

Most ``free'' services are advertisement supported. The less digitally literate user is under the impression that a possible service subscription is avoided by  consenting to the occasional addition of  random advertisements to the service. However, more the literate user knows that advertisements are targeted, their selection being based on the user's recorded behavior during interaction with the service or with other services, which sometimes are offered by other providers. Effectively it is the value of user's personal data that somehow balances a possible service subscription. Service subscriptions are not expected to exceed a few dollars per month. Popular paid services, with which a user could make comparisons, are in the order of magnitude of \$10; known examples are Netflix and Spotify. Social media typically avoid the subscription model, thus no direct comparisons can be made, but the users' expectations would probably be in the same order of magnitude. Effectively, a user's personal data are exchanged with services that are worth a few dollars.

This fact creates in the user's mind the impression that privacy itself is of low value. We shall see later that field research confirms this impression. On the other hand Google and Facebook  have continuously expanded their percentage in the technology sector stock market by squeezing the share of more traditional technology giants as IBM, Microsoft, Cisco and Apple \cite{bigsqueeze}. Apparently, gigantic amounts of monetary value are produced by offering ``free'' search and social networking services, entailing to a considerable extent privacy engineering. Furthermore,  a flourishing (albeit quietly working) personal data brokerage market exists. Effectively there seems to be a remarkably wide gap between (i)~the 
value attributed to personal data by  the data originators, who give the data  for free and~(ii) the revenues created by the aforementioned and other companies and organizations by using these data. The data using entities are effectively ``machines'' that convert personal and other data into unprecedented profits. 

Of course this is not the first time that information is monetized. However, personal information is not exactly  like any other type of information. Do the aforementioned considerations imply that service users (who sell privacy data low) are unfairly treated by service providers (who achieve high profits)? In some cases, as in YouTube, the company declares the exact amount of money that goes back to users as content creators, but social media users are not in a position to assess the balance between the profits of the social network provider and the value of the services offered to them.

\section{Personal data value estimation approaches}

\subsection{Market based approach}
The price of a  product or service is usually determined by a market, and value estimates are not independent of markets. Ideally we would find the value and price of a personal data set in a market. 
However, such a market between data creators and data users rarely exists. Suppose that I install in my computer a data logging software that records certain actions of mine and I decide to sell the log of one month; there is no market for such a transaction. 
Conversely, suppose that a company is willing to access my recorded behavior in a series of specific  transactions on a given service interface. 
They could make me an offer for this data set, which has been created by the interaction of their service with me, but they invariably ask for my consent to allow them to receive this data stream for free.

In certain situations  I could infer the profits of a company by using my personal data  from existing options. 
For example Spotify   offers ad free  music streaming for \euro5-14  per month (depending on the country). Alternatively, a free service is offered with ads, and weaker sequencing control of  songs on mobile phones. Effectively Spotify quits inserting advertisements and (maybe) collecting my data  for the above price. Although the pricing is certain to take into account the expected balance between paying and non-paying customers, the aforementioned price remains a strong indication of the personal data value. On the other hand Wolfram, whose highly sophisticated product Mathematica is priced in the thousands of dollars, is probably not concerned with user data.

There is yet another missing factor, namely the customer's quantified willingness to pay for a service, i.e. the service value as estimated by the customer. 
Suppose that a social network user is  offered a no-ads - no-tracking option for a paid subscription. How much would the average user be willing to pay for such an option? Most probably, taking into account the aforementioned price of subscriptions of streaming, gaming etc. services, a realistic  price for social networking services  is in the range $\char36 5-20$.


Yet another privacy value indication is likely to appear soon by  Google, which  has declared an intention to allow content publishers to charge for the content in the presence of ad blockers \cite{AdWords2017}. While this looks like a non-disturbance buy price, it will become a service sell price if publishers set the prices according to a content valuation scheme. Alternatively, they may try to set a price that will offset the loss due to  ad blockers, or to set a more or less flat ``fine'' that will deter  blocker usage. Notice that these are quite different valuation methodologies, and  that some of them may lead to irrational effects. Consider the following example: You start a video on a website, but before it starts an advertisement tries to pop up. If it is blocked, a window appears that asks for a payment. While you would expect the price to somehow reflect the value of the content, if it is just an offset to the expected advertisement gain (or even worse if it is a flat ``fine''), the connection between the real content value and the price that appears on the window is lost (due to the possibility to insert an ad before the video). The only connection that remains is that a more efficient advertisement is likely to appear together with a more highly valued video.

In view of the haziness in today's landscape, some researchers have proposed clearer arrangements. Perhaps we should pay a small amount each time we receive a service, and the service provider should pay us for our personal data sets or should return to us a part of their profits \cite{beresford2003location}. However, we shall not consider this category of models in this paper.

While in practice there is hardly a market to sell your own personal data, there is definitely a market for resellers of your data packed together with the data of millions of others.
Despite the numerous personal and sensitive data protection laws there is a market for personal data, brokers exist, buyers also exist and, obviously, there are initial sellers that harvest the data in the first place and sell them to the brokers, unless we assume that all data offered by resellers are products of security breaches.
According to Pasquale \cite{pasquale2014dark} 
personal data markets were already  a $\char36 156$-billion-a-year industry in 2014 in US, but the Federal Trade Commission (FTC) could find neither the data sources nor the final consumers. 
Not only the data are sold in a black underground market, but they are also often used for unlawful purposes.

In 2014 in a report of FTC \cite{federal2014data} the results of exploring the behavior of nine data brokers are outlined:
The data are  collected without consumers' knowledge; 
brokers buy data from different sources and exchange data with each other; the population coverage is almost complete and extensive per person; the data are analyzed and potentially sensitive inferences are made.

However the differential value of additional  data is likely to be strongly decreasing. Kosinski says that about 300 Likes on Facebook are enough to predict sensitive undisclosed traits of a person with an accuracy close to 100\% \cite{youyou2015computer}. Therefore the next few Likes will add nothing to a person's already complete picture. Since data brokers exchange, mix, and improve datasets, everybody's permanent picture will be complete in short time and  additional data will add zero information.

\subsection{Indirect privacy value  estimations}
Unfortunately, the informational privacy bears on other  types of privacy, i.e. on accessibility and on decisional privacy (\cite{tavani2011ethics}, see also section \ref{sect:introduction}). For some people a misuse of their personal data violates  their abstract notion of privacy, to which they attribute value without being willing to take steps to further elaborate on specific and concrete problems that are created or may be created by the aforementioned misuse. Persons of this type  may be reluctant to join social networks and may provide minimum personal information to online shopping sites; thus they are not willing to contribute to the creation and circulation of personal data. Perhaps their behavior provides an indication  that the value they attribute to privacy is higher than average. Is it possible to find out how these persons valuate their privacy?

According to Marion Fourcade if people mobilize and defend certain ``things,'' then these things must have value, even if they are not readily tradable or they do not appear in markets. Moreover, {\em the trade-offs that people make to pursue them are a good enough indication of the monetary value they implicitly attach to them}  \cite{fourcade2011price}. Fourcade gives an example of value estimation: What is the value of a pristine albeit difficult to access beach? It is higher than the value of our usual beach at least by an amount of money $v$, which is equal to 
the additional cost to travel to the better beach. The maximum such differential amount (paid by the most remote travelers) will give the additional value.

For example, if the average person in a given social environment is a member of certain  social networks, and uses a number of other services, which extract personal data, and if we somehow calculate the value of all these social networking and other services to an amount $V\!,$ then an abstainer that lives in the same environment values his or her privacy more than~$V\!.$

We might be tempted to use this methodology in order to calculate a value for our personal data despite the fact that we do not intend to sell them. One might say that the value of our personal data is equal to the maximum amount of money we are willing to pay for their protection. However, the situation is similar to an insurance related decision. The amount of money that we are willing to pay for the insurance of a house is not equal to the value of the house, but is rather in the order of magnitude of the house value times the probability that it will be destroyed.
Therefore in order to use this methodology we would have to know both possible damages and associated probabilities.
Actually this brings us closer to the damage base  approach, which we analyze in one of the forthcoming sections.


\subsection{Subjective  valuation}

As a matter of fact privacy champions are few, while the  personal data of the majority of the population are open to being harvested. 
There has been several studies that try to assess the value of personal data as estimated by their own producers, i.e. the persons involved in their creation.

Carrascal et al have conducted a study with an aim to estimate the value that users attached to their own personal data \cite{carrascal2013your} in 2013 in Spain. Users of online services were paid for their personal data according to a reverse second price auction. The study has found that users valued certain ``identity'' data about themselves (i.e. age, gender, address, economic status, thus permanent data) at about \euro 25, while they were willing to give away data collected during their browsing activity for only \euro 7. They could also ``sell'' their photos to social networks for \euro 12 and to search service providers for \euro 2. Effectively, they valued their personal information at about a few tens of US dollars, often much less.

In 2012 Beresford et al \cite{beresford2012unwillingness} have conducted a field experiment, in which they sold two similar DVDs online, but  they offered the second DVD for a somewhat reduced price (\euro 1 cheaper) while asking for the buyer's personal details. They found that  almost all users chose the cheaper product. When they set equal prices to both DVDs, buyers bought almost equal quantities, hence the title of their article ``unwillingness to pay for privacy.''

A meta-study created by Baruh et al \cite{baruh2017online}, i.e. a study that combines results from several previous studies, has concluded the following: (a) Users with enhanced privacy concerns are less likely to use online services, with the marked exception of social media (see next paragraph). (b) Concerned users share less information and adopt more privacy protective measures. (c) However,  digitally literate users tend to feel more comfortable with using online services. 

In another established trend in field studies, users of social media \cite{baruh2017online} are less concerned about privacy. They express personal opinions and upload content in public. The reduced privacy ``paradox'' that is associated with social media can be attributed mainly to self-fulfillment and self-promotion traits
. However the fact that users tolerate the harvesting of their behavioral and other data by the social network provider can also be attributed to the effectively monopolistic nature of social media like Facebook or Instagram. The {\em third person bias} has been offered as yet another explanatory argument by \cite{hbronlineprivacy}: Users tend to believe that 
the risks will only materialize on others.

\subsection{The damage based approach}
Damage based methodologies are popular in courts for the calculation of compensations, but they can also be used in a more general context. When a person's data end up in somebody else's possession, any control over their usage is lost. Given today's landscape, in which data are legally and illegally collected, sold, resold, intercepted, etc., and further used in uncontrolled and unforeseen ways, the damage that may result is extremely difficult to predict. Even past damages are difficult to assess.

To the  uncertainty in data dissemination one should in the near future add increased dangers due to smart objects. A person's medical record leakage could in the past have led to a rejection of a job application, but tomorrow the same knowledge combined with a hack of this person's smart medication tracking system may end up in a murder.

Anyhow, the value of  data  is often assessed by making reference to possible damage resulting by leakage, according to different scenarios and probabilities.
A~more general damage and compensation approach can also be applied.
For example, if targeted ads take an hour of one's time when using a social network app, the value of the resulting  disturbance could be set equal to one hour's average payment. As an example, suppose that a ``free'' audio streaming service inserts one 2 minute ad every 10 minutes. If a person is likely to use the service 5 hours per month, and if her hourly wage is $\char36 8$, then her monthly annoyance is equivalent to $\frac{2}{10} \times 8 \times 5 = 8 \, \char 36$. If the streaming service is priced to $\char36 7$ per month, both options are roughly equivalent.

However,  damage calculations are less convenient
(a)~when the damage is totally subjective, e.g. when she feels uncomfortable with her neighbors if they know her medical history, (b)~when the harm is done without the person's knowledge  \cite{Landwehr2016}, (c)~when the damage cannot be easily quantified, (d)~when the harm is done at a later time, and (e)~when  calculation is simply too complex.
Consider the scenario implied by articles \cite{grassegger2017data,anderson2017rise}, i.e.  alleged manipulation of voters in general elections by exploiting personal data harvested by social networks.
Possible damages  extend over an entire population,  are extremely difficult to calculate, and sufficient information may not be available before several decades. 
Last but not least in the same example, the damage could be purely ethical: Are we willing to bend the democracy rules, even if the data have been exploited by the ideal candidate?
The utilitarian answer to ethical, environmental and other intangible criteria is that the gain/loss calculation should be applied over a sufficiently broad context, i.e. given enough time and space, but this again ends up with a very complex calculation.
\subsection{The revenues approach}

In 2008 Enders et al. \cite{enders2008long}
surveyed a number of social networking sites and explored their revenue models. The top 5 sites were
MySpace, Classmates, Orkut, Friendster and Spoke, with 130, 40, 37, 36, and 35 million users respectively. Facebook users were only 13 million at the time (now more than two billion). All of them relied on advertising, while Classmates used subscriptions too.
Nine years later (in June 2017) the top five sites were Facebook, YouTube, Instagram, Twitter, and Reddit with 1940, 1000, 700, 313, and 250 million users respectively \cite{petekallas15mostpopular}. The videos of YouTube (which is owned by Google) are accompanied by advertisements, but 
55\% of the revenues are returned to video creators through the YouTube Partner Program. Effectively the YouTube model relies heavily on user creativity. Instagram also relies on advertisement since its acquisition by Facebook (2012). Twitter makes money through direct ads and by encouraging people to tweet about a brand for a reward. Reddit uses a complicated mixed model that relies on community support, advertisement and subscriptions. Reddit  claims that it does not track users.

The total revenue from advertisement that was gained by social networks in 2016 was about \$33 billion. Facebook's share was about 2/3 of the 33 billion. In 2017 Facebook alone is expected to reach 33 billion \cite{BakerTrends17,MediaBuying17}.
Facebook's average revenue per user (ARPU) was \$4.23   in the first quarter of  2017 \cite{Balakrishnan17,wiki:facebookpopulation2017}. A linear projection would give an annual ARPU of 
\$17,  but until the end of 2017 it will probably reach \$20. This is significant progress from an ARPU of around \$2 in 2009, and a significant growth since last year. About half of these revenues come from the mobile app.

What is the total value that a user brings to Facebook? If we calculate
 the total revenues during the  so far life of Facebook, these are close to \$80 billion, but they have been accumulated by a changing size of population. If we sum up the yearly ARPU in the last ten years, the result is about \$52; this is approximately the revenue brought by a ten year ``old'' Facebook member. However, a person's value for a social network is not equal to the value of the corresponding personal data.
 A social network makes money even from the mere membership of an otherwise inactive member,   due to the network externalities phenomenon. Size counts for social networks.
 
\subsection{Taking into account the  market value of vendors}
As we have said before, Google and Facebook have recently dominated the technology sector. However, Facebook is not a capital intensive company. Facebook owned network equipment equal 
to \$3.63 billion in 2015 \cite{fbequipment2015}. If the current population is taken into account we could estimate the current equipment to 4.5 billion.
Of course Facebook needs more than servers, which reside in huge data centers (or server farms).

However, the company's value in the stock market (market capitalization)
is around \$336 billion.
Effectively, the per user current value of Facebook is equal to 336/2 = \$168.  Even if we subtract a 10\% for ``hardware'' (that includes the real estate) and operating costs, we can estimate the intangible per user value of Facebook around \$150. Which part of this amount could be attributed to users' personal data is for further discussion.
One might further argue that market capitalization includes certain intangible factors too, e.g. shareholders' expectations for the future of a company. Anyhow, the amount of \$150 should be seen as a (perhaps not so tight) upper bound on the 
current user data value.

Note that Facebook's capitalization value is approximately equal to its valuation over a four year period, assuming that the valuation includes existing assets and future earning over a given time period. Such a valuation is usual when a company is about to enter the stock market.

\section{Privacy in social media}
In this section we explore the special relationship between privacy and social media.
Online social networking plays an exceptional role both in gathering data and in using data for producing revenues. Members willingly disclose personal and often sensitive information for the sake of participation and belonging.

\subsection{The privileged relationship between social networking and personal data collection} 

Privacy was unknown in traditional societies. Cohesion was an effective defensive weapon of small communities. A sharp division between {\em us} and {\em them} existed \cite{diamond2013world}. Modern cities brought privacy, but also dismantled community ties.
Social networks capitalize on such problems by partly restoring  ties. People effectively turn to the services of Facebook, LinkedIn, Google+, Instagram, Foursquare etc. and willingly reveal personal details, even secrets.  However, the  total online behavior of service users is  recorded. Personal data streams are generated, collected, sold, resold, exchanged, combined, and processed. 
Not only are the social media the best place to collect personal data, but any privacy concerns are also neglected by their users \cite{hbronlineprivacy}.

In addition, the results and conclusions that are generated are not limited to the obvious. Sky is the limit to the science of information processing and inference making. 
According to \cite{kosinski2013private,youyou2015computer} sensitive private traits can be uncovered  by examining a number of Likes on Facebook. Such powerful tools could open new prospects to marketers, extortionists, political parties, religious parties etc. Allegations for political manipulation  have already appeared in the press \cite{grassegger2017data}.

\subsection{The major revenue models}
Is personal data harvesting the only means of financially supporting a social networking service?
In fact the known  revenue tools in the area of social networking are (a)~advertising, (b)~subscription, and (c)~transaction models, which involve transaction fees, and  sales of products and/or services.
The latter can be generated by the network provider, by a third party, or by users. Both (a) and (c) require a critical mass of users \cite{enders2008long}. There is often a prevalent tool, which accounts for the greatest part of the revenues and defines the character of the network.
A  user's reaction to a subscription is different from a reaction to an one time fee. Service providers prefer longer time relationships with users, i.e. subscriptions.

There is discussion and criticism on the tendency of users towards ``free'' services and on the resulting  implications on the world of Internet. However,  the preference of service providers for subscription schemes (in telephony, in press, in magazines, in clubs etc.) has also played an important role, since they have taught the customer base that there is nothing between a free service and a monthly subscription service. 
Today's service management tools could possibly enable very short term subscriptions or even a fee per single transaction. For example, why should I be tied to a specific mobile telephony and data provider? 
Why can't I have an account with multiple (or all)  providers and choose the best service on a transaction basis? Service providers dislike any  increase in competition that will suppress prices, but customers should be more active in demanding alternative arrangements. There are indications of change. Streaming content providers now offer subscriptions with a ``cancel anytime'' option.

\subsection{The role of network externalities} 
In social networking an already grown social network will attract more new users than a young   one, since the existing user base largely determines the networking opportunities of new users. This phenomenon is known as {\em network externalities} or {\em network effects} \cite{shapiro1998information}. In 2014 {\em Ello} was created as an ad-free alternative to existing social networks. In the beginning they used an invitation policy and they gathered some momentum. However, it soon became clear to new members that their friends stayed back in the traditional large and ad infested social networks; the momentum was lost. Today Ello is not really a Facebook competitor but rather a specialized artists' network. 

By promoting a ``free'' social networking service model, Facebook and other networks are able to better exploit network externalities and get rid of competition in a possible market of different social networking services differently priced. Of course creating a ``free'' service by using advertisement is not  new, as it has extendedly been used for decades by traditional TV networks, and it is now used by the ``free press.''  The novelty lies in the extensive usage of personal data.

Consider a world of social networking sites that use a paid subscription model. A newcomer in this market would perhaps offer a free trial period, while waiting for the first generation of subscribers. 
Old established networks could charge more due to their increased user populations. In fact they would probably consider an optimized policy 
that would  take into account the increase of the population in the total revenue.

\subsection{Privacy policies in social networks}
Social networks are huge personal data collectors. Members willingly provide information hoping that in this manner they will improve their chances to network themselves with past or future friends. 

A privacy policy is implemented by using rules, which are described in a  document, but they must also be publicly accessible on a webpage and they must be written in easy to understand language. 
Since Facebook is the most often quoted example of information collecting and sharing network, we take some time to review their current privacy policy (June 2017). They state that they track their users from what they themselves write or upload, and from what other users write or upload. They say that they use this information to show their users targeted ads. They share user information within the Facebook family of companies, and they transfer information to third parties that support Facebook's operations by measuring the effectiveness of ads. They also give anonymized data to academic researchers. Apparently, this privacy policy does not allow Facebook to sell data to third parties.
Terms change from time to time in directions of reduced or increased privacy, trying to accommodate new interests or to make corrections after user protests. There is also a set of privacy settings, which allow for some user customization. However, Fuchs \cite{fuchs2012political} observes 
that privacy settings are available and work only against other users, not against advertisers.

\section{Other considerations}
\subsection{Upper bounds on value}
Even if a value for personal data could be determined by complex market mechanisms or other methodologies, certain external bounds are also applicable. For example, no food seller can expect to gain from me by analyzing my recorded behavior more than the amount of money I spend for food in a given time period. Effectively, the value of my (food related)  personal data would rather be a reasonably small percentage of my total food expenses. 

As a further example, consider the usage of social networking data in elections, in the spirit of the article of Grassegger and Krogerus \cite{grassegger2017data}. In this scenario a political party or their consultants ask for the personal data of a user of a social network. The network provider sells the data for a price, but a percentage  goes as a payment to the person that owns the data and has approved the exchange.  Again, it is difficult to determine a value for such an item, as it depends, among other factors, on the probability that the data will finally be converted to a favorable vote. And even if a vote does materialize, how critical is this vote for winning the election? Note also that there may exist alternative means towards the same end, and these alternatives imply price limits. For example, in a corrupt democracy certain politicians will directly buy votes, thus the vote street price will be an upper bound to the value of the personal data that may lead to winning a vote. 
\subsection{Transaction models and and data exchange}
Most users would probably answer that in principle they would benefit from increased privacy protection, despite the fact that in practice they give up too easily.
However Varian \cite{varian2002economic} has given certain examples on how
a potential buyer could improve the effectiveness of an exchange by giving more information to potential sellers. Assume that a person receives multiple messages from telecom operator agents trying to sell telephony subscriptions.  
The fact that the sales person is not fully aware of the preferences and the statistical profile of the phone calls of the client creates a discussion that is both annoying and time consuming. In other words, if the seller knew more about the client, then she could probably make a more targeted offer or would abstain from making an offer. In this model, the client is motivated to give up privacy in favor of both better services and less annoyance.

In fact this argument is not particularly original, as it is used every time we accept a first party cookie that is used by 
our service provider in order to serve us better without having to ask again about our  preferences. Since privacy is the right to be left alone, Varian's way is to include the lack of annoyance in privacy.

On the other hand, there are other types of information that we, as buyers, would not 
like to share with sellers. For example, we would keep the maximum price we would be willing to pay for a good secret. Effectively, w.r.t. to maximizing our benefit, including privacy aspects, we should be able to distinguish between information that we should give for free to the other side and information that we should keep for ourselves. Note also that the value of the aforementioned maximum price information is equal to the expected gain (for the seller) or loss (for the buyer) when this piece of information is used.

\section{Conclusions}
We number our main conclusions so as to facilitate a possible discussion:
\begin{enumerate}
\item There is a rough agreement, at least in the order of magnitude, between the subjective valuations of personal data and the earnings of large internet companies that monetize these data. For example, Facebook users might estimate the value of the service they receive, which is balanced by submitting their personal data, at a few dollars per month. On the other hand, today's annual average revenue per user (ARPU) for Facebook is equal to about \$20.
A superficial  conclusion that could be drawn  from the previous point might be that Facebook could perhaps produce more revenue by adopting a subscription model. However, it is rather obvious that a large part of its members would quit, thereby seriously damaging the network externalities on which Facebook is based.
\item The market capitalization of Facebook, Google, and similar companies, together with the fact that they are not capital intensive, point towards a value of more than \$100 per user, which however should be seen as accumulated over the time period of the existence of the company. Thus this amount is in rough agreement with the previous figures.
\item There is a complex data collecting system of companies and organizations, which is full of intentional and unintentional holes, as also shown by the existence of a data brokerage market. The end result of this evolution is that everybody's personal data will soon become common knowledge, at least among service providers, ending up with zero value. This tendency is reinforced by advances in data mining and reasoning, a glimpse of which appears in Kosinski's work \cite{kosinski2013private,youyou2015computer}.
\item Social media and search machine service providers have so far had a special relationship with personal data in  both collection and monetization, mainly through targeted advertising. However, in view of the dropping value of the (less volatile part of the) personal data, this sector 
is likely to (a)~enhance its role as a value extractor in the area of advertising, but also by expanding towards other areas (e.g. politics), (b)~focus on the extraction and exploitation of dynamic data (``where is John Doe, and what is he doing now?''), and (c)~control the dissemination of dynamic personal data, which will retain their value for a limited amount of time.
\item Needless to say, in this environment the average service user is not only less capable of protecting his or her own data, but also less likely to receive any compensation.
\item Any compensations (in the form of free services or payments)  are not likely to be designed to cover  probable damage from future data leakage and illegal exploitation. It is very difficult to predict the costs of such damages. This fact might create new opportunities for insurance companies.
\end{enumerate}

\bibliographystyle{alpha}
\bibliography{refspriv1}

\newcommand{\etalchar}[1]{$^{#1}$}
\begin{thebibliography}{EHDM08}

\bibitem[AH17]{anderson2017rise}
Berit Anderson and Brett Horvath.
\newblock The rise of the weaponized ai propaganda machine.
\newblock {\em Scout, February}, 12, 2017.

\bibitem[Bak]{BakerTrends17}
Dillon Baker.
\newblock The 6 most important social media trends of 2017.
\newblock [Online; accessed 22-June-2017].

\bibitem[Bal]{Balakrishnan17}
Anita Balakrishnan.
\newblock Facebook made about usd 4.23 off your profile in the first three
  months of the year.
\newblock [Online; accessed 21-June-2017].

\bibitem[BKP12]{beresford2012unwillingness}
Alastair~R Beresford, Dorothea K\"{u}bler, and S\"{o}ren Preibusch.
\newblock Unwillingness to pay for privacy: A field experiment.
\newblock {\em Economics Letters}, 117(1):25--27, 2012.

\bibitem[BS03]{beresford2003location}
Alastair~R Beresford and Frank Stajano.
\newblock Location privacy in pervasive computing.
\newblock {\em IEEE Pervasive computing}, 2(1):46--55, 2003.

\bibitem[BSC17]{baruh2017online}
Lemi Baruh, Ekin Secinti, and Zeynep Cemalcilar.
\newblock Online privacy concerns and privacy management: A meta-analytical
  review.
\newblock {\em Journal of Communication}, 67(1):26--53, 2017.

\bibitem[Buy]{MediaBuying17}
Media Buying.
\newblock Facebook mobile ad revenues to near \$30 billion next year.
\newblock [Online; accessed 22-June-2017].

\bibitem[CPN]{hbronlineprivacy}
Tomas Chamorro-Premuzic and Nathalie Nahai.
\newblock Why we?re so hypocritical about online privacy.
\newblock [Online; accessed 8-September-2017].

\bibitem[CRE{\etalchar{+}}13]{carrascal2013your}
Juan~Pablo Carrascal, Christopher Riederer, Vijay Erramilli, Mauro Cherubini,
  and Rodrigo de~Oliveira.
\newblock Your browsing behavior for a big mac: Economics of personal
  information online.
\newblock In {\em Proceedings of the 22nd international conference on World
  Wide Web}, pages 189--200. ACM, 2013.

\bibitem[{Dat}15]{fbequipment2015}
{Data Center Knowledge}.
\newblock The facebook data center faq, 2015.
\newblock [Online; accessed 14-June-2017].

\bibitem[Dia13]{diamond2013world}
Jared Diamond.
\newblock {\em The world until yesterday: What can we learn from traditional
  societies?}
\newblock Penguin, 2013.

\bibitem[eco]{bigsqueeze}
Cisco adapts to the rise of cloud computing.
\newblock {\em The Economist}.
\newblock [Online; accessed 4-July-2017].

\bibitem[EHDM08]{enders2008long}
Albrecht Enders, Harald Hungenberg, Hans-Peter Denker, and Sebastian Mauch.
\newblock The long tail of social networking.: Revenue models of social
  networking sites.
\newblock {\em European Management Journal}, 26(3):199--211, 2008.

\bibitem[Fou11]{fourcade2011price}
Marion Fourcade.
\newblock Price and prejudice: on economics, and the enchantment/disenchantment
  of nature.
\newblock {\em The Worth of Goods}, pages 41--62, 2011.

\bibitem[ftc14]{federal2014data}
Data brokers: A call for transparency and accountability.
\newblock Technical report, Federal Trade Commission, 05 2014.

\bibitem[Fuc12]{fuchs2012political}
Christian Fuchs.
\newblock The political economy of privacy on facebook.
\newblock {\em Television \& New Media}, 13(2):139--159, 2012.

\bibitem[GK17]{grassegger2017data}
Hannes Grassegger and Mikael Krogerus.
\newblock The data that turned the world upside down, 2017.

\bibitem[Kal17]{petekallas15mostpopular}
Pete Kallas.
\newblock Top 15 most popular social networking sites and apps, 6 2017.
\newblock [Online; accessed 15-June-2017].

\bibitem[KSG13]{kosinski2013private}
Michal Kosinski, David Stillwell, and Thore Graepel.
\newblock Private traits and attributes are predictable from digital records of
  human behavior.
\newblock {\em Proceedings of the National Academy of Sciences},
  110(15):5802--5805, 2013.

\bibitem[Lan16]{Landwehr2016}
Carl Landwehr.
\newblock Privacy research directions.
\newblock {\em Commun. ACM}, 59(2):29--31, January 2016.

\bibitem[Pas14]{pasquale2014dark}
Frank Pasquale.
\newblock The dark market for personal data.
\newblock {\em The New York Times}, 16:10--14, 2014.

\bibitem[Ram17]{AdWords2017}
Sridhar Ramaswamy.
\newblock Building a better web for everyone, 2017.

\bibitem[{Sta}17]{wiki:facebookpopulation2017}
{Statista}.
\newblock Number of monthly active facebook users worldwide as of 1st quarter
  2017 (in millions), 2017.
\newblock [Online; accessed 14-June-2017].

\bibitem[SV98]{shapiro1998information}
Carl Shapiro and Hal~R Varian.
\newblock {\em Information rules: a strategic guide to the network economy}.
\newblock Harvard Business Press, 1998.

\bibitem[Tav11]{tavani2011ethics}
Herman~T Tavani.
\newblock {\em Ethics and technology: Controversies, questions, and strategies
  for ethical computing}.
\newblock John Wiley \& Sons, 2011.

\bibitem[Var02]{varian2002economic}
Hal~R Varian.
\newblock Economic aspects of personal privacy.
\newblock In {\em Cyber Policy and Economics in an Internet Age}, pages
  127--137. Springer, 2002.

\bibitem[WB90]{warren1890right}
Samuel~D Warren and Louis~D Brandeis.
\newblock The right to privacy.
\newblock {\em Harvard law review}, pages 193--220, 1890.

\bibitem[YKS15]{youyou2015computer}
Wu~Youyou, Michal Kosinski, and David Stillwell.
\newblock Computer-based personality judgments are more accurate than those
  made by humans.
\newblock {\em Proceedings of the National Academy of Sciences},
  112(4):1036--1040, 2015.

\end{thebibliography}
\end{document}